\documentstyle[amssymb,preprint,aps,floats,psfig,prb,tighten]{revtex}
\newcommand{\Va}{NaV$_2$O$_5\;$}
\newcommand{\dg}{^\dagger}
\newcommand{\ve}{\varepsilon}
\newcommand{\k}{\kappa}

\begin{document}\draft \title{Superexchange in the quarter- filled\\
two-leg ladder system \Va} \author{V. Yushankhai$^{1,3}$ and P. Thalmeier$^2$}

\address{$^1$Max-Planck-Institute for the Physics of Complex Systems\\
$^2$Max-Planck-Institute for Chemical Physics of Solids\\ D-01187 Dresden, 
Germany\\
$^3$Joint Institute for Nuclear Research, 141980 Dubna, Russia} 
\date{\today} \maketitle 

\begin{abstract}

A theory of superexchange in the mixed valent layer compound \Va is presented
which provides a consistent description of exchange both in the disordered and
charge ordered state. Starting from results of band structure calculations 
for \Va first an underlying electronic model for a ladder unit in the
Trellis lattice is formulated. By using the molecular orbital representation
for intra-rung electronic states a second-order perturbation procedure is
developed and an effective spin-chain model for a ladder is derived.
Variation of the resulting superexchange integral $J$ is examined
numerically as the ladder system evolves from a charge disordered to the
extreme ('zig-zag') charge ordered state. It is found that the effective 
intra- ladder superexchange is always antiferromagnetic.

\end{abstract}

\vspace{1cm}

\newpage
\section{Introduction} 

Since the phase transition into a spin-gapped phase in $\alpha'$-\Va was 
reported
\cite{isobe}  this layered insulating compound has attracted much attention.
It was found that the phase transition ($T_c\simeq 34$ K) is accompanied
both by the lattice distortion \cite{chat} and a charge ordering 
\cite{ohama99}
in the vanadium layers. Though the details  of the low-temperature phase 
structure in \Va remain controversial, some main features of this structure
are obtained from experiment. For instance, the same low-T lattice 
superstructure was measured and reported by three groups of researchers
\cite{lud,boer,bern}. The experiment \cite{ohama,gren} 
has also ruled out a hypothetical 'in-line'
charge ordering (CO) in favour of a 'zig-zag' CO. At the same time 
mechanism of the spin-gap formation is not understood up to now and 
formulation of adequate spin models for \Va is required.

 The electronic and magnetic properties of \Va are mainly due to the V 
ions having  in average a mixed valence +4.5. In a particular layer
V-ions are arranged to give a sequence of two-leg quater-filled ladders
(one electron per rung) coupled to each other in a Trellis lattice. It is 
expected that  the main properties of vanadium layers are dominated by 
the ladder physics and the theoretical consideration should start
 with an analysis of this one-dimensional structure unit. The most
important evidence in favour of this point of view is provided by the band
structure calculations \cite{smol,yar}
within  the density-functional theory (DFT). These calculations have shown
that the overall electronic band structure of  \Va is determined mainly 
by the large intra-ladder electronic hopping amplitudes while the 
inter-ladder hopping is rather weak. 

Let us now turn to the problem of spin coupling in  \Va and consider a separate
quarter- filled ladder. Since the strongest electronic hopping transfers
$\sim$ t$_a$ between the basic vanadium d$_{xy}$ -orbitals are within the
rungs \cite{smol,yar} one has to ascribe a spin--$1/2$ not to 
a vanadium site but to the intra-rung V-O-V bonding molecular orbital. 
With strong Coulomb repulsion $\sim U$ on the d$_{xy}$ -orbital the lowest 
electronic bonding orbital cannot be doubly occupied and  the problem is
naturally reduced to a half-filled chain in the limit of strong electron
correlations. The excitation energy of the lowest 2-particle singlet rung
state from now on plays a role of an effective Hubbard repulsion parameter
which is roughly estimated to be 2t$_a$ in the charge disordered state 
\cite{smol,horsch}. Horsch and Mack were the first \cite{horsch} who derived 
a quantum 
Heisenberg spin--$1/2$ chain model to describe a one dimensional 
behavior of the ladder systems \Va in the charge disordered
 high-T phase. 

The superexchange  coupling parameters entering into the effective 
Heisenberg chain model are expected \cite{most,sa}  to be strongly 
reduced as the system undergoes the phase transition
into the low-T charge ordered  state. It was predicted, for instance,
that for neighbouring spins in a ladder the exchange integral $J$, 
which is
antiferromagnetic in high-T phase, may even change the sign to become a
ferromagnetic one in the extreme CO phase with a zig-zag charge
distribution on the ladder rungs \cite{sa}. In the present paper we 
develop the superexchange theory and obtain estimates for the spin-spin 
coupling constants in a single ladder both in the charge disordered and in an
ordered phase. That allows us to reexamine some earlier statements and
estimates  \cite{horsch,most,sa}. Since the in-line CO is ruled out by 
experiment  the present consideration is restricted to a zig-zag CO state 
only. 

The nature of the CO phase transition in the real \Va compound is not well
understood up to now. Nevertheless, at the first stage (Section III)
the analytical expressions for the superexchange integrals in this 
system can be derived  and analyzed in a rather general context.
Next, to reach a quantitative description one has to rely on a particular
realistic mechanism driving the charge ordering.  We assume that the
intersite Coulomb electron repulsion is responsible for the CO phase
transition and the lattice degrees of freedom play a secondary role. 
The inter-ladder short-range Coulomb interaction is treated in a mean
field approximation while intra-ladder charge-charge correlations are
considered on a more accurate level. To this end considering 
at some stage (Section IV) the charge degrees of freedom only we transform 
 the electronic Hamiltonian onto  the effective Ising model in the transverse
magnetic field \cite{thal,most}. The pseudospin variables in this approach are
associated with the charge dynamics and the order-disorder transition is
determined by the dimensionless interaction parameter
$\lambda=V_b/2t_a$. Here $V_b$ is the characteristic Coulomb repulsion
between electrons on neighbouring rungs. We adopt this approach which
allows us to examine quantitatively how the superexchange coupling varies
as the ladder system evolves from a disordered to the charge ordered
phase. We expect that besides the quantitative estimates  this particular 
analysis reveals general, mechanism-independent, trends in the behaviour 
of the superexchange spin-spin coupling in \Va.


\section{Electronic model and the eigenstate problem} 
We start with the electronic Hamiltomian for a quater-filled ladder in the
following general form

\begin{equation} 
H = \sum_{i}\left(H_0^{(i)} + H_{t,\perp}^{(i)}\right) +
\sum_{<ij>}\left(H^{(ij)}_{t,\parallel} + H^{(ij)}_{V}\right)
\label{a1} 
\end{equation}

where $(i)$ and $(ij)$ refer to a particular rung or to a pair of nearest
neighbouring ($nn$) rungs, respectively. The intra-rung interactions
involved into ($H_0^{(i)} + H_{t,\perp}^{(i)}$) are described by the
electronic hopping amplitude $t_a$, the Coulomb on-site, $U$, and
inter-site, $V_a$, parameters. For the $nn$ inter-rung coupling
($H_{t,\parallel}^{(ij)} + H_V^{(ij)}$), there are two, $t_b$ and $t_d$,
electronic hopping amplitudes and the corresponding Coulomb repulsion is
given by $V_b$ (see Fig.1). By introducing the creation operator
$d_{iL\sigma}^\dagger$ ($d_{iR\sigma}^\dagger$) for a spin- $\sigma$
electron on the left (right) V--ion within the i-th rung we specify the
particular terms in Eq.(\ref{a1}) as follows:

\begin{eqnarray} 
H_0^{(i)} &=&\varepsilon \sum_{\alpha=L,R}  n^\alpha_i + V_a n^L_i n^R_i +
U\sum_{\alpha=L,R} n^\alpha_{i\uparrow}n^\alpha_{i\downarrow}
\label{a2}\\
H^{(i)}_{t,\perp} &=& -t_a\sum_{\sigma}\left(d^\dagger_{iL\sigma}d_{iR\sigma} +
h.c.\right) \label{a3}\\
H_{t,\parallel}^{(ij)} &=& -t_b\sum_{\alpha=L,R}\sum_{\sigma}\left(d^\dagger_{i\alpha\sigma}d_{j\alpha\sigma} +
h.c.\right) \label{a4}\\
&&- t_d\sum_{\sigma}\left(d^\dagger_{iL\sigma}d_{jR\sigma} +d^\dagger_{iR\sigma}d_{jL\sigma}
+ h.c.\right)\nonumber\\
H^{(ij)}_V &=& V_b\sum_{\alpha=L,R} n^\alpha_i n^\alpha_j
\label{a6}
\end{eqnarray}
where $n_i^\alpha = \sum_\sigma n_{i\sigma}^\alpha = \sum_\sigma
d_{i\alpha\sigma} ^\dagger d_{i\alpha\sigma}$ ($\alpha=L,R$) and
$\varepsilon$ is a bare vanadium $d_{xy}$-orbital energy in an isolated
ladder.

Let us fix now the range of the model parameters. We use the following set
of hopping amplitudes \cite{yar}: $t_a$=0.38 eV, $t_b\simeq t_d$ =0.085
eV. The proper signs of these amplitudes are adopted in the definition of
$H$. Importance of additional hopping processes $\sim t_d$, which couple
$d_{xy}$- orbitals of two V-- ions at the opposite ends of  $nn$ rungs of a
ladder (Fig. 1), was reported in Ref.(\onlinecite{yar}). Approximate equality
$t_d\simeq t_b$ allows one to explain the pronounced feature of the
band structure in \Va. Namely, there is almost no dispersion of the
antibonding- type molecular orbital states formed by the $d_{xy}$--
orbitals on a rung. We will consider also another set: $t_a$=0.38 eV,
$t_b$=0.17 eV and $t_d$=0, as a more representative one among the sets
with $t_d$=0 used by other authors \cite{smol,horsch,most,sa}. 
The value of the on- site
Coulomb repulsion $U$ is fixed from the supercell LDA+U calculations
\cite{yar} to be $U=4$ eV. The estimates for the magnitudes of intersite
Coulomb repulsion $V_a$ and $V_b$ (Fig. 1) given in the literature are
less accurate. In Ref.(\onlinecite{sa}) for instance, it is claimed that a
screening reduces $V_a$ and $V_b$ parameters to rather small values
$V_a$, $V_b$ $<$ 0.4 eV, and hence the inter- rung Coulomb repulsion
cannot stabilize CO. To investigate this point we developed a combined
analysis by using the results of DFT calculation \cite{yar}
 and more generic multi-orbital
Hubbard model for a ladder in the Trelllis lattice. We found that
inter-orbital vanadium d-d electron correlations lead to an enhancement
for the effective $V_b$ parameter which enters into the effective
single-orbital Hamiltonian Eqs.(\ref{a1}--\ref{a6}). These results will 
be presented
elsewhere \cite{yush}. In the present investigation we take an intermediate
value for $V_a\simeq$0.5 meV and allow $V_b$ to vary in a wide range up to
1 eV.\\ It is worth noting that the validity of the perturbation procedure
developed below to the second order in the inter-rung hoppings is based
mainly on the smallness of these hoppings, $t_b$ and $t_d$, compared to
$2t_a$ provided the on-site Coulomb repulsion U is the largest parameter
of the theory. This guarantees also that the quarter- filled ladder is in
an insulating state \cite{smol,horsch,riera}. In contrast to the 
perturbation approach
developed in Ref.(\onlinecite{sa}) in our case there are no formal
restrictions on the values of $V_a$ and $V_b$. Like in
Ref.(\onlinecite{horsch}) these parameters could be arbitrary small giving a
meaningfull description  of the superexchange in a charge disordered state
of \Va. Surely, $V_b$ should be large enough, $V_b > V_b^{(\mbox{cr})}$, to
sustain a zig-zag CO state and the critical value $V_b^{(\mbox{cr})}$ depends 
on the mechanism chosen for the CO phase transition.\\

In the perturbation procedure the zeroth order Hamiltonian is defined as
$H_0=H-\sum_{<ij>}H_{t,\parallel}^{(ij)}$ where the last term is a
perturbation mixing electronic states on $nn$ rungs. As the necessary
prerequisite of the superexchange theory the 1- and 2- particle electronic
states on a particular rung have to be found explicitly. That can be done
by considering the inter- rung Coulomb interaction, the term
$\sum_{<ij>} H_V^{(ij)}$ in $H_0$, in a mean field (MF) approximation. In
this case the inter- rung Coulomb interactions lead to the following shift
of the on- site orbital energies $(\alpha = L,R)$:

\begin{equation} 
\varepsilon_{i\alpha}=\varepsilon + V_b\left[\langle
n^\alpha_{i+1}\rangle_g + \langle n^\alpha_{i-1}\rangle_g\right]
\label{a7}
\end{equation}
where $\langle n_{i\pm1}^\alpha\rangle_g$  are the averages over the lowest 
singly occupied electronic states of two rungs neighboring to the i-th rung
within the same ladder. Having $\sum_\alpha\langle n_i^\alpha\rangle_g=1$, 
one obtains
$(\varepsilon_{iL}+\varepsilon_{iR})/2=$$\varepsilon+V_b\equiv\bar{\varepsilon}$ and
\begin{equation}
\varepsilon_{iL}-\varepsilon_{iR}=\Delta\varepsilon_i=(-1)^i\Delta\varepsilon;
\qquad \Delta\varepsilon=2V_b\eta 
\label{a8}
\end{equation} 
In Eq.(\ref{a8}) we introduced the parameter $\eta$ ($0\leq\eta\leq 1$) of
charge disproportionation $\langle n_{iR}\rangle_g-\langle n_{iL}\rangle_g$$=(-1)^i\eta$ in a
zig-zag CO state of a ladder.\\

After applying the approximation given in Eq.(\ref{a7}) the zeroth order
Hamiltonian $H_0$ becomes a sum over individual rungs,
$H_0\simeq\sum_iH_0^{(i)}$. From now on each term $H_0^{(i)}$ can be
diagonalised independently. It is worth emphasizing that in this approach
the local intra- rung correlations due to $U$ and $V_a$ are treated exactly.
To specify  explicitly a form of $H_0^{(i)}$ in a particular $n-$
particle sector of the intra- rung electronic states we invoke an extra
subscript like in  $H_{0,n}^{(i)}$ below. For instance, by solving the
eigenstate problem in the 1-- particle sector one obtains
\begin{equation} 
H^{(i)}_{0,1}=\sum_{\sigma}\left[\varepsilon_g g^\dagger_{i\sigma} g_{i\sigma}
+ \varepsilon_f f^\dagger_{i\sigma} f_{i\sigma}\right] 
\label{a9}
\end{equation}
where the energy $\varepsilon_{g/f}$ of the lowest/excited molecular orbital
is given by
\begin{equation} 
\ve_{g/f}=\bar\ve\mp 1/2\sqrt{(\Delta\ve)^2 + (2t_a)^2}
\label{a10} 
\end{equation}
and
\begin{eqnarray} 
g\dg_{i\sigma}|0_i\rangle &=& \left[u_i d\dg_{iR\sigma} + v_i
d\dg_{iL\sigma}\right]|0_i\rangle \equiv |g_{i\sigma}\rangle\nonumber\\
f\dg_{i\sigma}|0_i\rangle &=& \left[-v_i d\dg_{iR\sigma} + u_i
d\dg_{iL\sigma}\right]|0_i\rangle \equiv |f_{i\sigma}\rangle
\label{a11}
\end{eqnarray}
The $u_i$ and $v_i$ coefficients in Eq.(\ref{a11}) are defined in the 
following way
\begin{equation} 
u_i=\sqrt{\frac{1}{2}\left[1 + \frac{\Delta\ve_i}{\sqrt{(\Delta\ve)^2 +
(2t_a)^2}}\right]}; \qquad  
v_i=\sqrt{\frac{1}{2}\left[1 - \frac{\Delta\ve_i}{\sqrt{(\Delta\ve)^2 +
(2t_a)^2}}\right]}
\label{a12}
\end{equation}
and for $\Delta\ve_i\neq 0$ the values of  $u_i$ and $v_i$ alternate
as i runs along the ladder. The complete basis for the 2- particle states
on the i-th rung consists of three singlets $|S_{i,n}>$
($n=1,2,3$) and triplet states $|\tau_i^{(m)}>$ ($m=0, \pm 1$). Explicitly, we
choose the following definitions for these states (the rung index i is
implied):
\begin{eqnarray} 
&&|S_1\rangle = 1/\sqrt{2}\left(d\dg_{R\uparrow} d\dg_{L\downarrow} -
d\dg_{R\downarrow} d\dg_{L\uparrow}\right)|0\rangle\nonumber\\ 
&&|S_{2,3}\rangle = 1/\sqrt{2}\left(\pm d\dg_{R\uparrow} d\dg_{R\downarrow} +
d\dg_{L\uparrow} d\dg_{L\downarrow}\right)|0\rangle\label{a13}\\
&&|\tau^{(0)}\rangle = 1/\sqrt{2}\left(d\dg_{R\uparrow} d\dg_{L\downarrow} +
d\dg_{R\downarrow} d\dg_{L\uparrow}\right)|0\rangle\nonumber\\
&&|\tau^{(\pm 1)}\rangle = d\dg_{R\uparrow} d\dg_{L\uparrow}|0\rangle, \quad 
d\dg_{R\downarrow} d\dg_{L\downarrow}|0\rangle\nonumber
\end{eqnarray}
One can check that the triplets are degenerate eigenstates of
$H_{0,2}^{(i)}$, i.e.
$H_{0,2}^{(i)}|\tau_i^{(m)}\rangle=\varepsilon_\tau|\tau_i^{(m)}\rangle$, 
with the energy $\varepsilon_\tau$$=2\bar{\varepsilon}+V_a$. 
In the singlet subspace  the
Hamiltonian $H_{0,2}^{(i)}$ takes the following form
\begin{eqnarray} 
H^{(i)}_{0.2} = \left(|S_{i,1}\rangle,|S_{i,2}\rangle,|S_{i,3}\rangle,\right)
\left(\begin{array}{lll}
2\bar\ve + V_a  &-2t_a  &0\\
-2t_a           &2\bar\ve + U  &\Delta\ve_i\\
0               &\Delta\ve_i &2\bar\ve + U
\end{array}\right)
\left(\begin{array}{l}
\langle S_{i,1}|,\\
\langle S_{i,2}|,\\
\langle S_{i,3}|,
\end{array}\right)
\label{a14} 
\end{eqnarray}
By diagonalizing the symmetric 3$\times$3 matrix one obtains the
eigenstates
\begin{equation} 
|S_{i}^{(\k)}\rangle = \sum_{n=1}^{3}\gamma^{(\k)}_n(i)|S_{i,n}\rangle
\label{a15} 
\end{equation}
and the corresponding eigenvalues $\varepsilon_{s\kappa}$ ($\kappa$=1,2,3) in
the singlet subspace. 
More precisely,  
$\varepsilon_{s\kappa}=2\bar\varepsilon+V_a+(U-V_a)e_\kappa$,
where $e_\kappa  (\kappa=1,2,3)$, are the solutions of a cubic 
characteristic equation reading
\begin{eqnarray} 
e_\kappa\left[(1-e_\kappa)^2-\zeta_i^2\right] +\mu^2[1-e_\kappa]=0
\label{15a}
\end{eqnarray}
where $\zeta_i=\Delta\varepsilon_i/(U-V_a)$ and $\mu=2t_a/(U-V_a)$.
Note, that according to Eq.(\ref{a8}) the value of $\zeta_i^2$ does not depend
on $i$. Provided  Eq.(\ref{15a}) is solved the 3-component vector
$\vec\gamma^{(\kappa)}(i)$ in Eq.(\ref{a15}) can be found by using the
following expressions
\begin{eqnarray} 
\gamma_1^{(\kappa)}&=&\pm\frac{|A_\kappa|}{R_\kappa}\nonumber\\
\gamma_2^{(\kappa)}&=&\pm sign(A_\kappa)\frac{\mu(1-e_\kappa)}{R_\kappa}\\
\gamma_3^{(\kappa)}(i)&=&\mp \label{15b}
sign(A_\kappa)\frac{\mu\zeta_i}{R_\kappa}\nonumber\label{15b} 
\end{eqnarray}
where $A_\kappa=(1-e_\kappa)^2-\zeta_i^2$ and
$R_\kappa=[A_\kappa^2+\mu^2$$(1-e_\kappa)^2+\mu^2\zeta_i^2]^{1/2}$.
Note, that the upper/or the lower signs in Eq.(\ref{15b}) should be taken 
simultaneously.
Like $\zeta_i$, the third component $\gamma_3^{\kappa}(i)$ is staggered along
the chain direction.
This particular problem, Eqs.(\ref{15a}, \ref{15b}),
can be solved numerically in a wide range of the underlying model
parameters provided the order parameter $\eta$ is fixed from some
self-consistent procedure. Most representative results will be discussed
in Sec.V.\\

For further purposes it is helpful to use the projection operators
$X_i^{q,q}=|q_i\rangle\langle q_i|$ and the transfer operators
$X_i^{q,p}=|q_i\rangle\langle p_i|$
where  $|q_i\rangle$ and $|p_i\rangle$ are the states defined above on the 
i-- th rung.
More explicitely, for instance, one has
\begin{eqnarray} 
X_i^{g\sigma, g\sigma} &=& |g_{i\sigma}\rangle\langle g_{i\sigma}|, \qquad  
X_i^{0, g\sigma} = |0_{i}\rangle\langle g_{i\sigma}|,\label{a16}\\
X_i^{S\k, S\k} &=& |S_{i}^{(\k)}\rangle\langle S_{i}^{(\k)}|, \quad 
X_i^{S\k, g\sigma} = |S_{i}^{(\k)}\rangle\langle g_{i\sigma}|
\quad \mbox{etc.}\nonumber
\end{eqnarray}
With these notations the approximate zeroth order Hamiltonian
$H_0\simeq\sum_i H_0^{(i)}$ can be written as
\begin{eqnarray}
H_0^{(i)} &=& \ve_v X_i^{0,0} + \ve_g\sum_{\sigma}X_i^{g\sigma, g\sigma} +
\ve_f\sum_{\sigma}X_i^{f\sigma, f\sigma} + \nonumber\\
&& + \sum_{\k=1}^3 \ve_{s\k}X_i^{S\k, S\k} +
\ve_\tau\sum_{m=\pm 1,0} X_i^{\tau m, \tau m} \label{a17} 
\end{eqnarray}
In Eq.(\ref{a17}) the vacuum rung state $|0_i\rangle$ with the reference energy
$\varepsilon_v$ is involved for completeness. 

At the same time the 3-- and 4-- particle states are dropped as these 
states are irrelevant for the present
purposes. At the end of this section the following remark is worth to be
made. Actually, while solving the intra- rung eigenstate problems we could
not avoid the use of a MF treatment of inter- rung Coulomb interactions,
the term $\sum_{<ij>}H_V^{(ij)}$ in the original Hamiltonian of Eq.(\ref{a6}).
Nevertheless it does not mean the complete neglect of the many body
effects due to $H_V^{(ij)}$ in our approach. Further analysis in Secs.
IV,V will show that charge-charge correlations between singly occupied $nn$-
rungs are important to obtain the estimates for the superexchange coupling
between spins in a ladder.


\section{Perturbation procedure and\\ superexchange on a ladder} 
The
superexchange coupling between two singly occupied rungs is established
due to the electron hopping transitions with the virtual intermediate
2-particle states in one of two rungs. These transitions are due to the
kinetic part of the Hamiltonian, H$_{t,\parallel}$, whose action has now to
be represented in terms of the molecular orbital states $|q_i\rangle$ 
defined in the previous section. In a rather general form one may write:
\begin{eqnarray} 
H_{t,\parallel} = \sum_{<ij>} H^{(ij)}_{t,\parallel} = \sum_{<ij>, \sigma}
\sum_{qq^\prime} \sum_{pp^\prime} \left[t_{ij, \sigma} (q^\prime,
q|p^\prime, p) X^{q^\prime, q}_{i} X^{p^\prime, p}_{j} + h.c.\right]
\label{a18}
\end{eqnarray} 
where
\begin{eqnarray} 
t_{ij, \sigma} (q^\prime, q|p^\prime, p) &=& -t_b\sum_{\alpha}\langle
q^\prime_i|d\dg_{i\alpha\sigma}|q_i\rangle\langle
p^\prime_j|d_{j\alpha\sigma}|p_j\rangle  \nonumber\\
&& - t_d\left[\langle q^\prime_i|d\dg_{iR\sigma}|q_i\rangle\langle
p^\prime_j|d_{jL\sigma}|p_j\rangle + (R \leftrightarrow L)\right]
\label{a19} 
\end{eqnarray}
and ($R\leftrightarrow L$) means the indices interchange. All the necessary
amplitudes $\langle q'_i|d_{i\alpha\sigma}^\dagger|q_i\rangle$ and their 
conjugates can
be now straightforwardly evaluated to give the relevant part of
$H_{t,\parallel}$ in the following form
\begin{eqnarray} 
H_{t,\parallel} &=& -\sum_{ij,\sigma}
\xi_{ij}\left\{\frac{(2\sigma)}{\sqrt{2}}\sum_{\k=1}^{3}
D^{(\k)}_{ij}\left(X_i^{S\k, g\bar\sigma} X_j^{0, g\sigma} + h.c.\right)
+\right.\nonumber\\
&&\left. + \sum_{\sigma^\prime}\sum_{m=\pm 1,0}
B_{ij} C^{(m)}_{\sigma^\prime\sigma}\left(X_i^{\tau m, g\sigma^\prime} X_j^{0, g\sigma} + h.c.\right)
\right\}\label{a20} 
\end{eqnarray}
Here $\xi_{ij}=1$ if $(ij)$ is a $nn$ pair and equals zero otherwise;
$(2\sigma)=+1(\sigma=\uparrow)$, $-1(\sigma=\downarrow)$ and
$\bar{\sigma}=-\sigma$. The factor $C_{\sigma'\sigma}^{(m)}$ is defined as
\begin{equation} 
C^{(m)}_{\sigma^\prime\sigma} = \delta_{\sigma^\prime, \sigma} \delta_{m,
2\sigma} + 1/\sqrt{2} \delta_{\sigma^\prime, \bar\sigma} \delta_{m,0}.
\label{a21} 
\end{equation}
The effective transfer amplitudes $D_{ij}^{(\kappa)}$ and $B_{ij}$ can be
presented in the following form:
\begin{eqnarray} 
D^{(\k)}_{ij} &=& \sum_{n=1}^3 \gamma^{(\k)}_n(i)\left[t_b a_{ij,n} + t_d \bar
a_{ij,n}\right]\nonumber\\
B_{ij} &=& t_b b_{ij} + t_d \bar b_{ij} \label{a22}
\end{eqnarray}
where $a_{ij, n}$, $\bar{a}_{ij, n}$, $b_{ij}$ and $\bar{b}_{ij}$ are
bilinear functions of the $u$--, $v$-- coefficients
\begin{eqnarray}  
&&a_{ij,1} = \bar a_{ij,2} = u_i v_j + v_i u_j \nonumber\\
&&a_{ij,2} = \bar a_{ij,1} = u_i u_j + v_i v_j \label{a23}\\
&&a_{ij,3} = \bar b_{ij} = -u_i u_j + v_i v_j \nonumber\\
&&b_{ij} = \bar a_{ij,3} = -u_i v_j + v_i u_j \nonumber
\end{eqnarray}
To derive the superexchange spin coupling to the second order in the
effective hopping amplitudes $D_{ij}^{(\kappa)}$ and $B_{ij}$ we use the
Schrieffer-Wolff transformation $\tilde{H}$=$\exp(-\hat{S})H\exp(\hat{S})$ 
with the
generator $\hat{S} (= -\hat{S}^\dagger)$ being determined from the condition
$[H_0,\hat{S}]=-H_{t,\parallel}$. Then the second order correction to $H_0$ is
given by
\begin{equation} 
H_{\mbox{eff}} = -1/2 \left[\hat{S}, H_{t, \parallel}\right] \label{a24}
\end{equation}
From Eq.(\ref{a20}) we obtain first the form of the generator
\begin{eqnarray} 
\hat{S} &=& \sum_{ij, \sigma} \xi_{ij}\left\{\frac{(2\sigma)}{\sqrt{2}} \sum_{\k=1}^3
\frac{D^{(\k)}_{ij}}{\ve_{s\k}+\ve_v -2\ve_g} X^{S\k, g\bar\sigma}_i X^{0,
g\sigma}_j + \right.\nonumber\\
&& \left. +\sum_{\sigma^\prime, m}\frac{B_{ij}
C^{(m)}_{\sigma'\sigma}}{\ve_{\tau}+\ve_v -2\ve_g} X^{\tau m, g\sigma'}_i X^{0,
g\sigma}_j - h.c. \right\}
\label{25}  
\end{eqnarray}
One can see, that there are two perturbation channels, $(s)$ and $(\tau)$,
with the singlet and triplet intermediate virtual states, respectively.
Finally we obtain from Eq.(\ref{a24}) the superexchange Hamiltonian
\begin{equation}
H_{\mbox{eff}} = \sum_{<ij>}\left(J^{(s)}_{ij} + J^{(\tau)}_{ij}\right) \vec
S_i \vec S_j 
\label{a26}
\end{equation}
with the exchange constants in the singlet and triplet channel given by
\begin{eqnarray}
J_{ij}^{(s)} &=& \sum_{\k=1}^3\frac{[D_{ij}^{(\k)}]^2 +
[D_{ji}^{(\k)}]^2}{\ve_{s\k} + \ve_v - 2\ve_g} \label{a27}\\
J^{(\tau)}_{ij} &=& - \frac{B^2_{ij} + B^2_{ji}}{\ve_\tau + \ve_v -2\ve_g}
\label{a28}  
\end{eqnarray}
In Eq.(\ref{a26}) the spin- $\frac{1}{2}$ operators
$S_i^\alpha=1/2\sum_{\sigma\sigma'}|g_{i\sigma}\rangle\sigma_{\sigma\sigma'}^\alpha\langle
g_{i\sigma'}|$
and $\sigma_{\sigma\sigma'}^\alpha$ are the Pauli matrices
$(\alpha=x,y,z)$; the notation $<ij>$ means that each pair of the $nn$
rungs is taken only once.\\

Note that already at this stage the expressions Eq.(\ref{a27}) and 
Eq.(\ref{a28})
together with the definitions in Eqs. (\ref{a22}), (\ref{a23}), 
formulate a complete scheme to estimate $J^{(s)}_{(ij)}$ and
$J^{(\tau)}_{(ij)}$ both in a disordered and a zig-zag CO state. That can
be done, at least numerically, by solving first the eigenstate problems of
Sec.II for given parameters of the model and with the order parameter
$\eta$ varying in the range $0\leq\eta\leq 1$. In this approach the effects
of charge ordering upon spin coupling in a ladder can be examined in its
main features without specifying a mechanism driving the CO transition in
a quarter- filled ladder system. Preliminary analysis shows that the total
superexchange coupling $J_{(ij)}=J^{(s)}_{(ij)}+J^{(\tau)}_{(ij)}$
is reduced as the ladder develops from a disorderd to a zig-zag CO state.
That is in accordance with a predicition in Ref.(\onlinecite{most}). A more
satisfactory quantitative description could be done if one relies on a
specific mechanism responsible for CO phase transition. In the next
section we pursue one possibility by noting that the underlying model,
Eqs.(\ref{a1} - \ref{a6}) contains the expected instability without invoking
extra degrees
of freedom like charge-phonon coupling. Further analysis is based on the
pseudospin description of the charge correlations \cite{thal,most}. 
Before turning to the pseudospin model itself (Sec.IV) we use this 
representation to present the main results of
this section given in Eqs.(\ref{a22}, \ref{a23},
\ref{a27}, \ref{a28}) in a more condensed form.

In the pseudospin representation the molecular ground state of the i--th
rung $|g_{i\sigma}\rangle=|g_i\rangle\otimes|\sigma_i\rangle$ is given by
\begin{equation} 
|g_{i}\rangle = \left[u_i|\uparrow\rangle_i + v_i|\downarrow\rangle_i\right];
\qquad u^2_i + v^2_i =1 \label{a29}
\end{equation}
By using the pseudo- spin ($T=\frac{1}{2}$) operators
$T_i^\alpha (\alpha=x,y,z)$ we define now several pair correlation
functions and expectation values in the ground state of a ladder as follows
\begin{eqnarray} 
Q^{(\parallel)}_{ij} &=& \langle T^z_i T^z_j\rangle_g, 
\qquad\quad Q^{(\perp)}_{ij} =
\langle T^x_i T^x_j + T^y_i T^y_j\rangle_g \nonumber\\
R^{(x)}_{ij} &=& \langle T^x_i + T^x_j\rangle_g, \qquad R^{(z)}_{ij} =
\langle T^z_i - T^z_j \rangle_g \label{a30}\\
R^{(xz)}_{ij} &=& \langle T^x_i T^z_j - T^z_i T^x_j\rangle_g \nonumber
\end{eqnarray}

 By using the above definitions and Eqs.(\ref{a22}, \ref{a23}) one may write 
the numerators
in the expressions of Eqs.(\ref{a27}, \ref{a28}) in the following form:
\begin{eqnarray} 
&&\left[D^{(\k)}_{ij}\right]^2 + (i \leftrightarrow j) =
4t^2_b\left\{\sum^3_{n=1}\left[\gamma^{(\k)}_n\right]^2 
{\mathcal{K}}^{(n)}_{ij}
+ 2\gamma^{(\k)}_1 \gamma^{(\k)}_2 {\mathcal{K}}^{(12)}_{ij} +
\right.\nonumber\\
&& + \left.  \gamma^{(\k)}_1\left[ \gamma^{(\k)}_3(i) -
\gamma^{(\k)}_3(j)\right]{\mathcal{K}}^{(13)}_{ij} +  \gamma^{(\k)}_2\left[
\gamma^{(\k)}_3(i) -  \gamma^{(\k)}_3(j)\right]{\mathcal{K}}^{(23)}_{ij}
\right\}
\label{a31}\\
&&B^2_{ij} + (i \leftrightarrow j) = 4t^2_b {\mathcal{K}}^{(4)}_{ij}
\label{a32}
\end{eqnarray}
where 
\begin{eqnarray} 
&&{\mathcal{K}}^{(1),(2)}_{ij} = \left(1 + \nu^2\right)\left[\frac{1}{4} +
Q^{(\perp)}_{ij}\right] \mp \left(1 - \nu^2\right) Q^{(\parallel)}_{ij} + 2\nu
R^{(x)}_{ij}\nonumber\\
&&{\mathcal{K}}^{(3),(4)}_{ij} = \left(1 + \nu^2\right)\left[\frac{1}{4} -
Q^{(\perp)}_{ij}\right] \pm \left(1 - \nu^2\right) Q^{(\parallel)}_{ij}\nonumber\\
&&{\mathcal{K}}^{(12)}_{ij} = 2\nu\left[\frac{1}{4} +
Q^{(\perp)}_{ij}\right] + \left(1 + \nu^2\right) R^{(x)}_{ij}\label{a33}\\
&&{\mathcal{K}}^{(13)}_{ij} = \nu^2 R^{(xz)}_{ij} - \frac{1}{2}\nu R^{(z)}_{ij}\nonumber\\
&&{\mathcal{K}}^{(23)}_{ij} = \nu R^{(xz)}_{ij} - \frac{1}{2}\nu^2 R^{(z)}_{ij}\nonumber
\end{eqnarray}
and  we defined $\nu=t_d/t_b$.
It is worth noting that the antisymmetric property,
${\mathcal{K}}_{ij}^{(13)}=-{\mathcal{K}}_{ji}^{(13)}$ 
and ${\mathcal{K}}_{ij}^{(23)}$$=-{\mathcal{K}}_{ji}^{(23)}$, makes the
complete expression of Eq.(\ref{a31}) to be symmetric.


\section{Pseudo- spin model for the ladder} 

The effective Hamiltonian in Eq.(\ref{a26}) describes spin-spin interactions 
in the low- energy subspace spanned by the vector charge- spin manifold
$|G_{\{g\},\{\sigma\}}\rangle$$=\prod_i[|g_i\rangle\otimes|\sigma_i\rangle]$. 
Our aim now is to project the electronic model, Eqs.(\ref{a1}-\ref{a6}) 
onto the charge sector within this manifold. The projection procedure was 
discussed several times \cite{thal,most}, it results in the following 
zeroth- order effective Hamiltonian for a ladder

\begin{eqnarray}
H^{(0)}_{\mbox{eff}} &=& -2t_a\sum_i T^x_i + 2V_b\sum_{<ij>}\left[T^z_i T^z_j +
1/4\right] - \nonumber\\
&& - \Delta\ve^0\sum_i (-1)^i T^z_i 
\label{a34}
\end{eqnarray}
In this one- dimensional (1D) model the effective longitudinal field
$\sim\Delta\varepsilon^0$ expresses in  a condensed form effects of Coulomb
interactions due to higher dimensionality of the real \Va compound (see
Appendix).

We proceed further in the spirit of the coupled quantum spin chain
approach \cite{scal} which involves a MF treatment of the inter-chain
coupling and a more sophisticated treatment of the intra-chain, i.e.
intra-ladder, interactions. The intra- ladder problem enters into the
expressions for the superexchange constants, Eqs.(\ref{a27}, \ref{a28}) 
and Eqs.(\ref{a30}-\ref{a33}), in the
form of pair correlation functions for the $nn$ pseudospins. Having the
final aim of calculating these constants we rely on properties of the
exact solution for the 1D Ising model in a transverse field.

The term $\sim\Delta\varepsilon^0$ in Eq.(\ref{a34}) breaks explicitly
 the Z(2) symmetry 
of the 1D transverse Ising model which results in a non-zero longitudinal
magnetisation $\langle T_i^z\rangle_g=(-1)^i\eta/2$ if the coupling constant
$\lambda=V_b/2t_a$ exceeds some critical value, $\lambda >\lambda_c$. One
may write
\begin{equation} 
\eta = \left[1 - \left(\frac{\lambda_c}{\lambda}\right)^2\right]^\beta
\label{a35}
\end{equation}
and choose for the exponent $\beta = 1/8$ since the presence of a weak
longitudinal field does not change the universality class of the 1D Ising
model in the transverse field. Note, that for this exactly solvable model
$\lambda_c=1$ (Ref. \onlinecite{lieb,pfeuty}) and we keep this value in the 
subsequent numerical
analysis. The open problem that still remains are unknown dependencies of
$\beta$ and $\lambda_c$ on details of the inter- ladder Coulomb
interaction if one goes beyond a mean field treatment of these interaction
in the Trellis lattice.

Let us now turn to a discussion of the superexchange constants,
Eqs.(\ref{a27}, \ref{a28}).
We estimate the inter- rung pseudospin correlation functions entering into
Eq.(\ref{a30}) at two different levels of sophistication. The simplest one 
is to decouple the pair correlation functions as follows:
\begin{eqnarray}
Q^{(\parallel)}_{ij} &\simeq& -\langle T^z_i\rangle^2 = -\left(\eta/2\right)^2\nonumber\\
Q^{(\perp)}_{ij} &\simeq& \langle T^x_i\rangle^2 =
1/4\left(1 - \eta^2\right)\label{a36}\\
R^{(xz)}_{ij} &\simeq& -\langle T^z_i - T^z_j\rangle\langle T^x_i\rangle =
(-1)^{i+1}\frac{\eta}{2}\sqrt{1 - \eta^2}\nonumber
\end{eqnarray}
In Eq.(\ref{a36}) we take into account that $i$ and $j$ are refered to $nn$ 
rungs and,
hence, $\langle T_j^z\rangle= -\langle T_i^z\rangle$ while $\langle T_j^x\rangle=\sqrt{1-\eta^2}/2$ does
not depend ont the rung index $i$. Within this first approximation the
superexchange constants will be calculated in the next section. The only
signatures of the exact solution in Eqs.(\ref{a35}, \ref{a36}) are due
to the choice of the
values for $\beta$ and $\lambda_c$, while the inter- rung fluctuations are
ignored. To gain an impression how these fluctuations may change the
superexchange constants we adopt some more properties of the exact
solution for the 1D transverse Ising model. For this purpose let us
consider \cite{lieb,pfeuty} the pair correlation functions $(j=i\pm1)$:
\begin{eqnarray} 
\langle T^z_i T^z_j\rangle_g = k_z; \quad \langle T^x_i T^x_j\rangle_g = m_x^2
+ k_x; \quad \langle T^y_i T^y_j\rangle_g = k_y \label{a37}
\end{eqnarray}
In particular, one has \cite{lieb,pfeuty}
\begin{eqnarray} 
k_z = -\frac{1}{4\pi}\int^\pi_0 dq \frac{\lambda + \cos(q)}{\Lambda_q}; \qquad
m_x = \frac{1}{2\pi}\int^\pi_0 dq \frac{1 + \lambda\cos(q)}{\Lambda_q} =
\langle T^x_i\rangle_g
\label{a38}
\end{eqnarray} 
where $\Lambda_q=\sqrt{1+\lambda^2+2\lambda\cos q}$. The expressions for
$k_{x,y}$ can be found in Refs.(\onlinecite{lieb,pfeuty}) as well. We use the
property $|k_x|,|k_y|\ll m_x,|k_z|$ and drop the weak transverse
fluctuations, which means k$_{x,y}\simeq$ 0. Breaking of the Z(2) symmetry
allows us to write also
\begin{equation} 
R_{ij}^{(xz)} \simeq (-1)^{i+1}\eta m_x 
\label{a40}
\end{equation}
We use the Eqs.(\ref{a35}, \ref{a37}) and (\ref{a38}) together with the
approximation in Eq.(\ref{a40}) as the basis of
our second approach to calculate the superexchange constants.


\section{Analysis of superexchange integrals,\\ numerical results} 

The starting point of our analysis is the exact diagonalisation of the 
single-rung Hamiltonian in the 1- and 2-particle sectors. The excited 
singlet and triplet states provide the intermediate states for the 
superexchange mechanism of magnetic coupling between neighbouring 
singly occupied rungs. From now on we drop the irrelevant lower indices
at $J$.
The excitation energies 
$E_{s\k}=\varepsilon_{s\k}+\varepsilon_{v}-2\varepsilon_g$ and
$E_{\tau}=\varepsilon_{\tau}+\varepsilon_{v}-2\varepsilon_g$ 
which enter as the denominators into the perturbative expressions for
$J^{(s)}$, $J^{(\tau)}$, are determined  by the eigenvalues due 
to Eqs.(\ref{a10}) and (\ref{15a}). These eigenvalues  depend on the degree 
of charge order via the intra-rung d- orbital levels shifts
$\Delta\varepsilon$ which is proportional to the CO parameter $\eta$ 
in Eq.(8). 
The symmetry breaking source is assumed to be arbitrary small. The excitation
energies in the  singlet and triplet channels
as function of $\Delta\varepsilon$ are shown in Fig.(2), $\ve_v =0$.
There is a crossing of two singlets at very high values of
$\Delta\varepsilon_i$. For realistic values of V$_b$ however this crossing
point is not reached. 

 While solving the eigenvalue problem in the singlet sector the components 
of the vectors $\vec{\gamma}^{(\kappa)}(i)$, Eq.(\ref{15b}), are calculated 
as well.
These vector components together with pseudospin correlation functions
determine the numerator, Eq.(\ref{a31}), in the expression for $J^{(s)}$,
Eq.(\ref{a27}). The pseudospin pair correlation functions are calculated 
either in
the modified MF approximation, Eq.(\ref{a36}), or by using a more accurate
approach based on Eqs.(\ref{a37})-(\ref{a40}). In the latter case the 
most pronounced 
charge fluctuation contributions to the superexchange are involved. This
requires  evaluation of only two pair correlation functions, namely,
$\langle T^z_i T^z_{i+1}\rangle_g = k_z$ and
$\langle T^x_i T^x_{i+1}\rangle_g \simeq m_x^2$
which are given by Eq.(\ref{a37}),(\ref{a38}). Both of them are shown
in Fig.(3) as function of $\lambda^{-1}$=$2t_a/V_b$. It is seen 
that the intra- chain fluctuations of the order parameter described by 
the 'longitudinal' correlation function
k$_z$ extend far above the critical value $\lambda_c^{-1}$=1 where the CO 
vanishes.
Since in our approximation k$_x \simeq 0$  the 'transverse'
correlation function is reduced to the square of the 'covalency' of
each rung $m_x=\langle T_i^x\rangle_g$. The latter approaches one half for
$\lambda^{-1}\rightarrow\infty$. It shows the opposite behaviour compared to
the complementary charge order parameter 
$m_z=|\langle T_i^z\rangle_g|=\eta/2$. The latter is
shown  in Fig.(3) as given by Eq.(\ref{a35}) with the exact 1D
Ising exponent $\beta$=$1/8$.

\noindent With these necessary ingredients it is now straightforward to
calculate the superexchange integrals $J^{(s)}$, $J^{(\tau)}$ as
function of the microscopic model parameters according to Eqs.(\ref{a27}),
(\ref{a28}). 
The hopping and interaction parameters  fixed from LDA+U- calculations are
discussed in Sec. II where definite values for $t_a, t_b, t_d$ and U
were given. The intra- rung repulsion $V_a$ was assumed to lie close to
0.5 eV and the inter- rung repulsion $V_b$ is allowed to vary in a wide
range up to 1 eV. The dimensionless control parameter for CO is
$\lambda$=$V_b/2t_a$ with $\lambda_c$=1 at the CO transition.\\ 
The dependence of $J^{(s)}$, $J^{(\tau)}$ on $V_b$ is shown in
Fig.(4) for two cases: The dashed line corresponds to exchange integrals
calculated in the modified MF approach, Eq.(\ref{a36}), with the Ising
exponent $\beta$=$1/8$. 
For $\lambda < 1$ ($V_b< V_b^{(\mbox{cr.})}=2t_a = 0.76$ eV) in the
disordered ladder the exchange constants are independent of $V_b$ and the
total value $J = J^{(s)} + J^{(\tau)}$ amounts to $J_{\mbox{\tiny MF}} =
J_{\mbox{\tiny MF}}^{(s)}\simeq 80$ meV since $J^{(\tau)}_{\mbox{\tiny MF}} =0$ in this region.
At $V_b = V_b^{(\mbox{cr})}$ a sharp drop in the singlet exchange constant 
is observed whose shape is determined by the order parameter $m_z$ 
(Fig.(3)). The most
important reason for the reduction of superexchange is the reduced
effective hopping between adjacent rung bonding states caused by charge
order. The full line in Fig.(4) shows the same exchange integrals with
the effect of charge fluctuations included, as described by the pseudo-
spin formalism of the previous chapter, Eqs.(\ref{a37})-(\ref{a40}). In the 
disordered regime 
the inclusion of charge fluctuations strongly reduces the exchange
constants compared to the MF value. The total superexchange integral
falls now into the region 70 meV$>J> 55$ meV, which is somewhat lower than the
theoretical value obtained in Ref.(\onlinecite{horsch}) for the disodered 
phase while it is somewhat higher than the value
$J^{(\mbox{exp})}\simeq50$ meV estimated in Ref.(\onlinecite{isobe}).
Fig.4  shows that in the CO regime
($V_b>V_b^{(\mbox{cr})}$), where charge fluctuations are strongly suppressed, 
both
approximations give very similar results. Above the critical region at, 
for instance,
$V_b\simeq 1.3V_b^{(\mbox{cr})}$ (i.e. at $V_b =1$ eV), where CO is nearly
complete, $m_z \simeq 0.43$, the superexchange integral reaches the value
$J\simeq 0.22$ meV.\\  
In Fig.(5) we show  $J^{(s)}$
and $J^{(\tau)}$as function of V$_b$, with charge fluctuation effects
included, for various intermediate values of the
intra-rung Coulomb repulsion $V_a$. 
Moreover, by varying $V_a$ in a wider range from zero up to rather high 
values, $V_a\sim 4t_a$, a regular behaviour of the exchange constants
is found, which is already seen from Fig.(5). First, 
the triplet contribution $J^{(\tau)}$ is almost negligible as compared to
the singlet part $J^{(s)}$ for the model with $t_d=t_b$. In the case $t_d$
=0 however $J^{(\tau)}$ gives a sizeable (negative) contribution to the 
total $J$. Secondly, with increasing $V_a$, the total
superexchange integral $J$ decreases and remains antiferromagnetic ($> 0$),
 it never changes its sign. This is true for both models considered
($t_d=t_b$ or $t_d=0$).
These observations are at strong variance with the results of 
Ref.(\onlinecite{sa})

  It is instructive to examine also both the limit of the extreme CO, 
$m_z\to 1/2$, which is reached at $\lambda^{-1}\to 0$,
i.e. at unphysicaly high values of $V_b$, and 
the the opposite limit  $\lambda^{-1}\to\infty$, i.e. at $V_b=0$.
 To check these limits a variation
of the total superexchange integral $J$ is examined  in a wide range of
$\lambda^{-1}$, Fig.6(a). Two curves there correspond to two sets of the
hopping parameters used: (1) $t_a=0.38$eV, $t_b=t_d=0.085$eV and 
(2) $t_a=0.38$eV, $t_b=0.17$eV and $t_d=0$. The second set is the most 
representative one among the sets with $t_d=0$ used by other authors
\cite{smol,horsch,most,sa}. First, at $\lambda^{-1}\to 0$ the curve (1)
reaches the
value $J=7.2$meV which coincide exactly with the estimate based on the 
standard expression $J=4t_d^2/U$. The latter estimate is due to the
superexchange coupling between two vanadium ions located at the opposite ends
of $nn$ rungs. From Fig.6 one can see that in the case of finite $t_d$
the superexchange is regularly enhanced as compared to the case with $t_d=0$.
In fact we found numericaly that if $\lambda^{-1}\to\infty$,
the two dependences, i.e. the curves (1) and (2) in Fig.6,
approach the same value $J=80$meV which is nothing than the MF value of
$J$ depicted in Fig.4. In this limit ($V_b=0$) the general formulas, 
Eqs.(\ref{a27}), (\ref{a28}),
can be easily treated analytically to give the following expression for the
total superexchange integral 
\begin{eqnarray}
J = 2t^2\left\{\frac{\left(1 + \mu\right)^2}{1 + \mu^2} \cdot \frac{1}{2t_a -
J_a + V_a} + \frac{\left(1 - \mu\right)^2}{1 + \mu^2}\cdot\frac{1}{2t_a + J_a
+ U}\right\}
\label{a41}
\end{eqnarray}
where $\mu=2t_a/(U-V_a)$; $J_a=(2t_a)^2/(U-V_a)$ and $t=t_b + t_d$
for the set (1) or $t=t_b$ for the set (2). By using 
the same parameter values as in Fig.6 one  finds that Eq.(\ref{a41})
reproduces completely the above mentioned numerical value for $J$.
A large change in $J$ seen in Fig.6(a) at $\lambda=\lambda_c$ is
provoked by a sharp behaviour of the order parameter $m_z$ as the system
undergoes the CO phase transition. Considering now the CO regime only  
($\lambda^{-1} < 1$)  we reproduce in Fig.6(b) the same variation of
$J$, however, with respect to the order parameter $m_z$. Fig.6(b) shows
a rather smooth behaviour with almost linear decrease of $J$ in the regime
of weak charge disproportionation, $m_z < 0.2$.


\section{Discussion and Conclusion}

In Sec.II we formulated the underlying electronic model for the compound \Va 
with most of the model paremeters ($t_a, t_b, t_d$ and U)
fixed from known results of the DFT calculations. By using this model
in Sec.III we developed the superexchange theory and derived the second-order
perturbation expression (in the effective V-V hopping amplitudes $t_b$ and
$t_d$) for the intra-ladder exchange integral $J$ both in the charge
disordered and in the zig-zag charge ordered state. Assuming that the Coulomb 
mechanism drives the charge ordering (Sec.IV) the exchange integral $J$
is calculated in a wide range of the variable $V_b$ and $V_a$ parameters.
Since qualitative results are presented in details in Sec.V we summarize
here most important qualitative findings.

\begin{enumerate}
\item[(i)] 
In the intra-ladder superexchange mechanism the triplet 
channel is of minor importance for the model with $t_d=t_b$ and the 
resulting exchange integral is dominated by the singlet channel, 
$J\simeq J^{(s)}$ (Figs.4,5). For the model 
with $t_d=0$, $J^{(\tau)}$ also gives a sizeable (negative) 
contribution to $J$.

\item[(ii)] 
In the charge disordered state inter-rung charge-charge 
correlations within a ladder  cause a strong reduction
of the exchange integral value (Fig.4).

\item[(iii)] 
With increasing of the zig-zag charge ordering the exchange integral
$J$ reduces strongly but never changes sign, i.e. $J$ remains 
antiferromagnetic ($>0$) for the whole parameter range, Figs.4-6.

\item[(iv)] 
Variation of the underlying electronic model due to different choices
of the hopping amplitude $t_d$ (either $t_d\neq 0$ or $t_d=0$) may lead
to a considerable quantitative change in the value of $J$, especially in the 
extreme CO limit where $J$ stays finite or vanishes respectively (Fig.6).
\end{enumerate}

 In common, Figs.4-6  reveal a general trend. That is the decrease of the 
total exchange integral $J$ as the inter-rung charge-charge correlations
within a ladder increase. We believe this trend together with less
obvious properties (i)-(iv) are of general significance not restricted to
the particular mechanism we used to describe the  CO phase transition in 
the ladder system.

 In the present paper we concentrated on a study of the the intra-ladder 
superexchange in \Va. To derive a more complete spin model in the
Trellis lattice a thorough analysis of the inter-ladder magnetic 
interactions is required. Such an analysis for the charge disodered
phase  in \Va was done in Ref.(\onlinecite{horsch}). There the authors 
claimed that due
to the almost perfect cancellation of the triplet- and singlet-interactions
in a-direction the ladders are effectively decoupled to give essentially
1D magnetic behavior of the entire spin system. In support of this conjecture
we recall also that in the charge disordered Trellis lattice the inter-ladder
Heisenberg spin-spin interactions are strongly frustrated. These arguments,
however, cannot be transfered directly to a zig-zag CO state. The DFT
calculation \cite{yar} of the exchange integrals in a zig-zag CO state 
have shown, for instance, a rather considerable ferromagnetic (FM)
inter-ladder coupling \cite{yar}.
The opposite sign of this coupling is not surprising in view of different 
symmetry of V-O-V bonds connecting neighbouring ladders. It is a 
a challenging problem to reveal the microscopical origins and estimate
a value of the resulting FM coupling between neighbouring ladders, which
can be done in a properly developed superexchange theory taking into
account also effects of direct V-V exchange.\\

\section*{Acknowledgement}
We are grateful to A. Bernert, P. Fulde, A. Yaresko and A.A. Zvyagin 
for numerous fruitful discussions.
One of the autors (V.Yu.) thanks the INTAS organization (INTAS-97-11066)
for support.


\newpage
\appendix

\section*{}

In this Appendix we exploit an extended model including inter-rung Coulomb 
interactions. Our aim is to show  that these interactions lead to the symmetry 
breaking term $\sim\Delta\varepsilon^0$ in Eq.(\ref{a34}).\\

We choose the notation $V_{ab}$ for the Coulomb repulsion between the
electrons on the diagonal V-V bond, i.e. on a pair of $nn$ V-ions belonging
to adjacent ladders (Fig.(1)). We assume the parameter $V_{ab}$ to be not
too large otherwise the instability with respect to the in- line charge
ordering would be the dominant one. Now the necessary addition to the single
ladder model, Eq.(\ref{a1}), can be presented in a  form
$H_{ab}=\sum_{\{ij\}}H_{ab}^{\{ij\}}$, where the summation is over the pairs
$\{ij\}$ of $nn$ rungs belonging to adjacent ladders. For a particular $\{ij\}$
pair one has
\begin{equation}
H_{ab}^{\{ij\}}= V_{ab}(i_L,j_R) n^L_i n^R_j =  V_{ab}(i_R,j_L) n^R_i n^L_j
\label{aa1}
\end{equation}
where the composite index $i_L$ ($i_R$) means the position of V-ion on the
left (right) side of the $i$-th rung; the definitions for the electron density 
operators $n_i^\alpha (\alpha=L,R)$ are the same as in the main text. In the
pseudospin representation the term $H_{ab}$ takes the following form
\begin{eqnarray}
H_{ab} &=& -\sum_i\Delta\ve^0_i T^z_i +
\sum_{\{ij\}} V_{ab}(i_{L}, j_{R}) \left(\frac{1}{4} - T^z_i T^z_j\right) =\nonumber\\
&& = H_{ab,1} + H_{ab,2}\label{aa2}
\end{eqnarray}
where
\begin{eqnarray}
\Delta\ve^0_i = \frac{1}{2}\sum_{j\in i}\left[
V_{ab}(i_{L}, j_{R}) - V_{ab}(i_{R}, j_{L})\right]\label{aa3}
\end{eqnarray}
and summation in (\ref{aa3}) runs  over the  rungs neighbouring to the 
$i$--th rung.\\

Without distortion in the Trellis lattice all the diagonal bonds are identical,
$V_{ab}(i_{L}, j_{R}) = V_{ab}(i_{R}, j_{L})= V_{ab}(|\vec{r_0}|)$. 
In this case
 $\Delta\varepsilon_{i}^0 = 0$ and the inter-ladder coupling is given by the
bilinear term $H_{ab,2}$. Assuming a zig-zag charge ordering in the ladders
one may check that in the mean field approximation this term is averaged
to zero due to a special geometry of the ladder stacking. It means that on
the mean field level the ladders in the non-distorted Trellis lattice are 
disconnected.
Below we argue briefly that  a symmetry breaking source appears if the 
lattice symmetry is lowered as reported, 
for instance, in Refs.(\onlinecite{lud},\onlinecite{bern}) for \Va.

L\"udecke et al. \cite{lud} proposed for the low- T lattice superstructure of
\Va two species of ladders, denoted as A and B in Fig.1.
 In each A- ladder
there is a weak transverse, along the a-axis, modulation of the rung
positions while the lattice structure of the B- ladders remain symmetric
as in the high-T phase. The doubling of unit cell means, in
particular, that in the A-B-A'-B .. array the rung's position
modulation in the A- and A'-ladders are of opposite signs. In the
modulated Trellis lattice a shortening/lengthening ($\pm\vec{\delta}$) of
the diagonal V-V bonds leads to a special spatial pattern for the intersite 
Coulomb parameter variation $V_{ab}(|\vec{r_0}\mp\vec{\delta}|)\simeq$
$V_{ab}(|\vec{r_0}|)\mp\nabla V_{ab}\vec{\delta}\equiv V_{ab}^{(\mp)}$ (see
Fig.1). By using Eq.(\ref{aa3})
one can easily check that this
pattern produces in the A-ladders 
an effective longitudinal field which is staggered along the ladder
b-direction,
$\Delta\varepsilon_{i,A}^0=(-1)^i\Delta\varepsilon_A^0$, with the
amplitude $\Delta\varepsilon_A^0\simeq 2|\nabla V_{ab}\vec{\delta}|$.
Moreover, the field $\Delta\varepsilon_{i,A}^0$ is also staggered along
a-direction, which sustains the zig-zag CO state with the A- and A'- ladders
being ordered in the antiphase manner.
Due to  smallness of the lattice distortion, 
$|\vec{\delta}|/|\vec{r_0}|\approx 10^{-2}$, we expect that
$\Delta\varepsilon_A^0 \ll V_b$ for the actual values of $V_b$ providing
the CO phase transition.\\
It follows from Eq.(\ref{aa3}) that a symmetry breaking field is not present 
explicitly in the B-subsystem, i.e. $\Delta\varepsilon_B^0=0$, for
the spatial pattern in Fig.1.  Strictly speaking, a presence of non-zero
field $\Delta\varepsilon_B^0$ is not necessary since the long-range zig-zag
CO in B-ladders, if it is present, can be explained due to
{\it spontaneous} symmetry breaking if $\lambda > \lambda_c$.
 
Note that besides the extracted staggered field term the
inter-ladder bilinear interaction, $H_{ab,2}$,  is still present. With a
zig-zag ordering in A-/or both in A- and B- subsystems this term being
considered in the MF- approximation is averaged to zero even in the modulated
Trellis lattice structure.\\

\newpage

\begin{figure}
\vspace{5cm}
\centerline{\psfig{figure=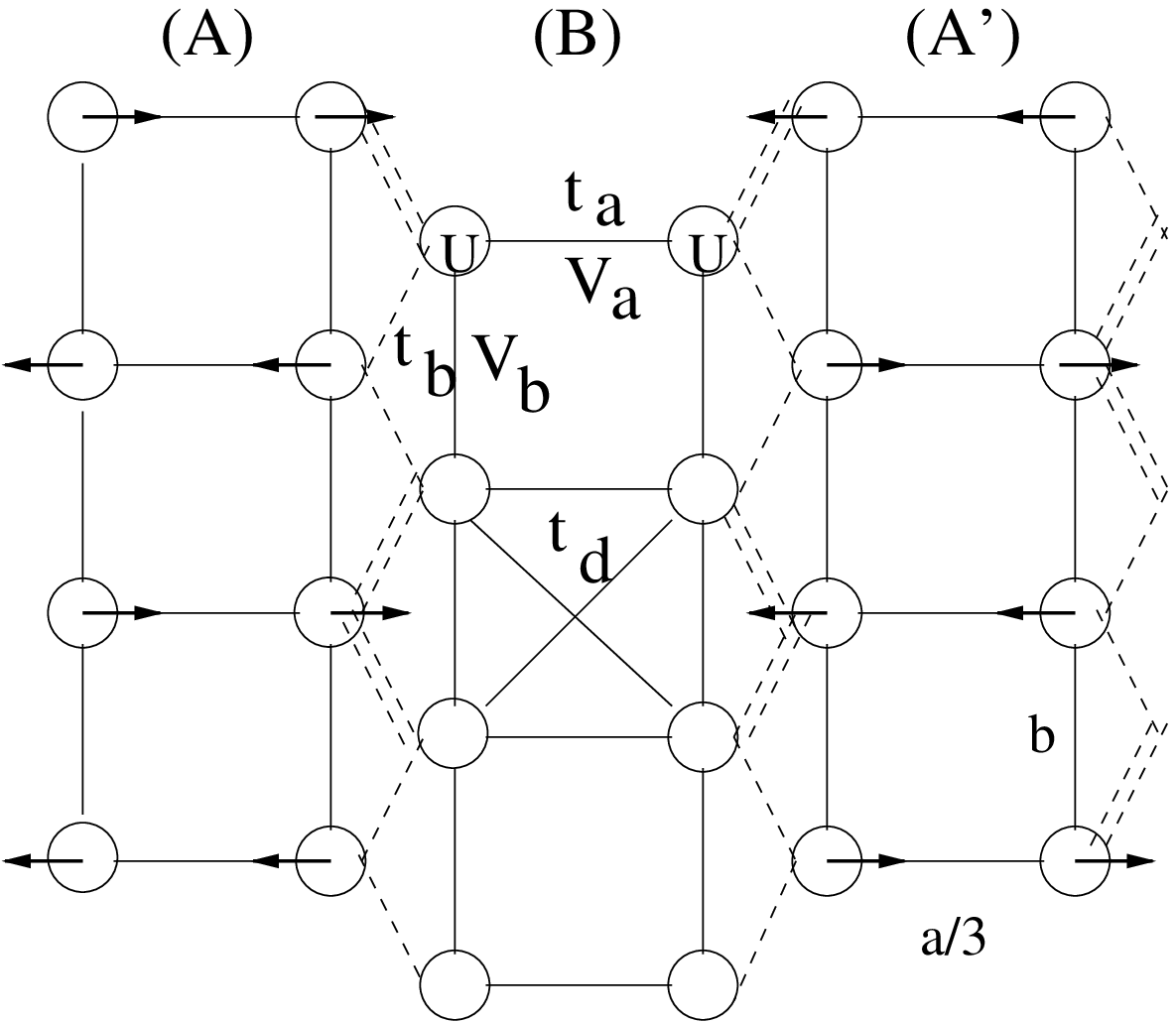,height=8cm,width=8cm}}
\vspace{1cm}
\caption{
Schematic representation of the crystal structure of a vanadium layer
of \Va. The vanadium ions are denoted by open circles, the hopping
amplitudes ($t_a, t_b$ and $t_d$) and Coulomb repulsion parameters ($V_a$,
$V_b$) are ascribed to the intra-ladder bonds; the on-site Coulomb repulsion
$U$ is also shown. Arrows indicate the weak modulation (not in scale) of 
V-ion positions in the low-T phase. The single (double) dashed lines are 
associated with $V_{ab}^{(+)}$ ($V_{ab}^{(-)}$) which are the inter-ladder
Coulomb repulsion parameters modulated in the low-T phase.}
\vspace{5cm}
\end{figure}

\begin{figure}
\centerline{\psfig{figure=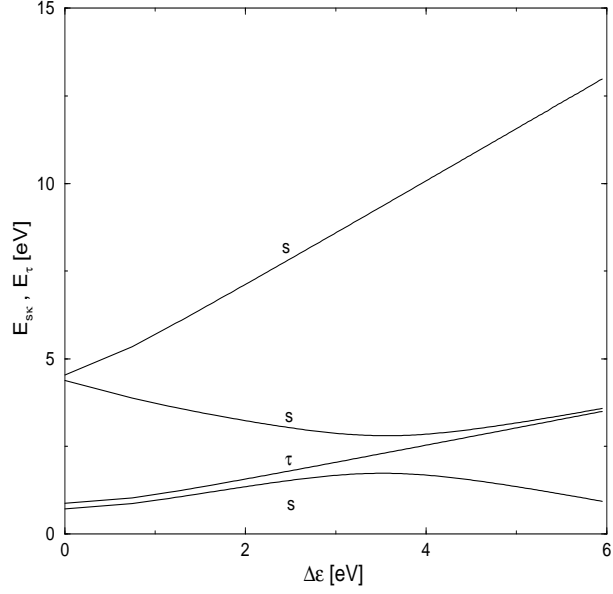,height=8cm,width=8cm,angle=-90}}
\vspace{1cm}
\caption{
The excitation energies in the singlet ($s$) and the triplet ($\tau$)
channels as functions of $\Delta\varepsilon$. The model parameters used
are (in eV): $t_a=0.38, t_b=t_d=0.085, U=4, V_a=0.5$.}
\end{figure}

\begin{figure}
\centerline{\psfig{figure=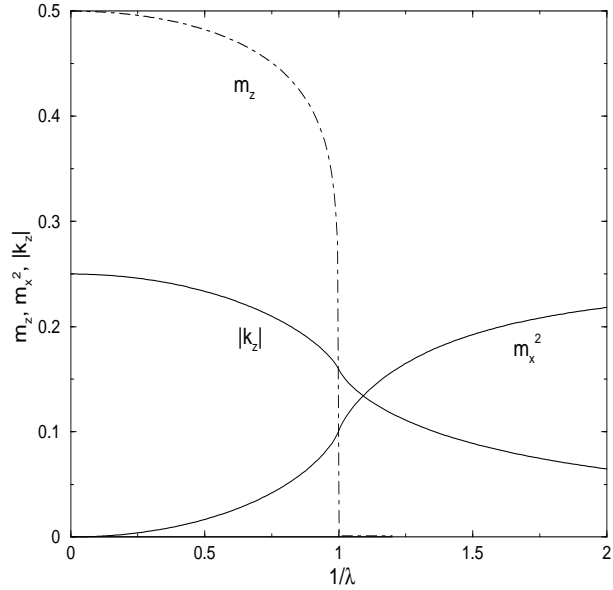,height=8cm,width=8cm,angle=-90}}
\vspace{1cm}
\caption{
Pseudospin correlation functions
$\langle T^z_i T^z_{i+1}\rangle_g = k_z$ and
$\langle T^x_i T^x_{i+1}\rangle_g \simeq m_x^2$
as functions of $\lambda^{-1}$=$2t_a/V_b$.
The order parameter $m_z=|\langle T_i^z\rangle_g|=\eta/2$
calculated with the exact 1D Ising exponent $\beta=1/8$ is shown with 
a dashed line.}
\end{figure}

\begin{figure}
\centerline{\psfig{figure=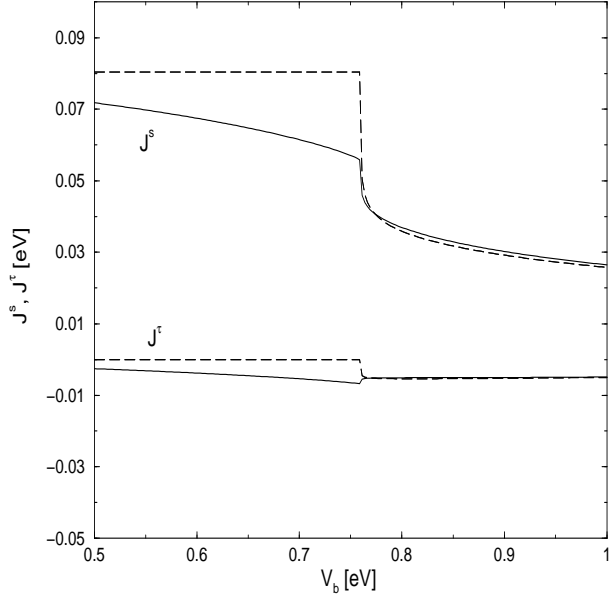,height=8cm,width=8cm,angle=-90}}
\vspace{1cm}
\caption{
Variation of the exchange integral $J^{(s)}$ and $J^{(\tau)}$ with respect
to $V_b$ (the other model parameters are the same as in Fig.1 and 
$\beta=1/8$). The integrals are calculated either in the modified MF approach
(dashed lines) or with the effect of charge fluctuations included (solid
lines).}
\end{figure}

\begin{figure}
\centerline{\psfig{figure=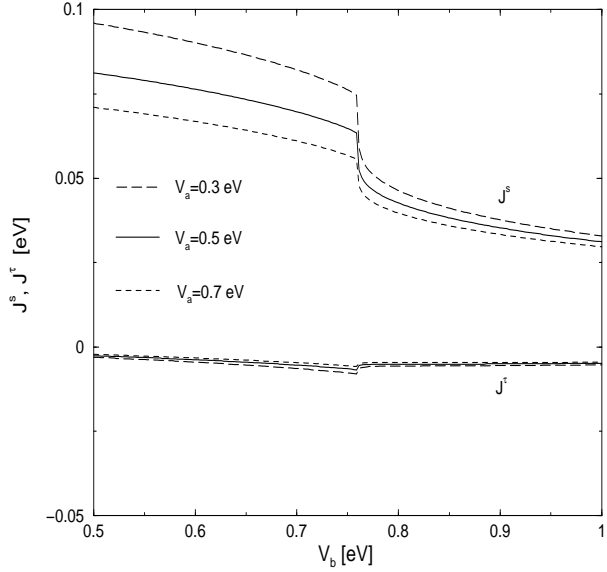,height=8cm,width=8cm,angle=-90}}
\vspace{1cm}
\caption{
The exchange integrals  $J^{(s)}$ and $J^{(\tau)}$
as functions of $V_b$ for various values of $V_a$.}
\end{figure}

\begin{figure}
\centerline{\psfig{figure=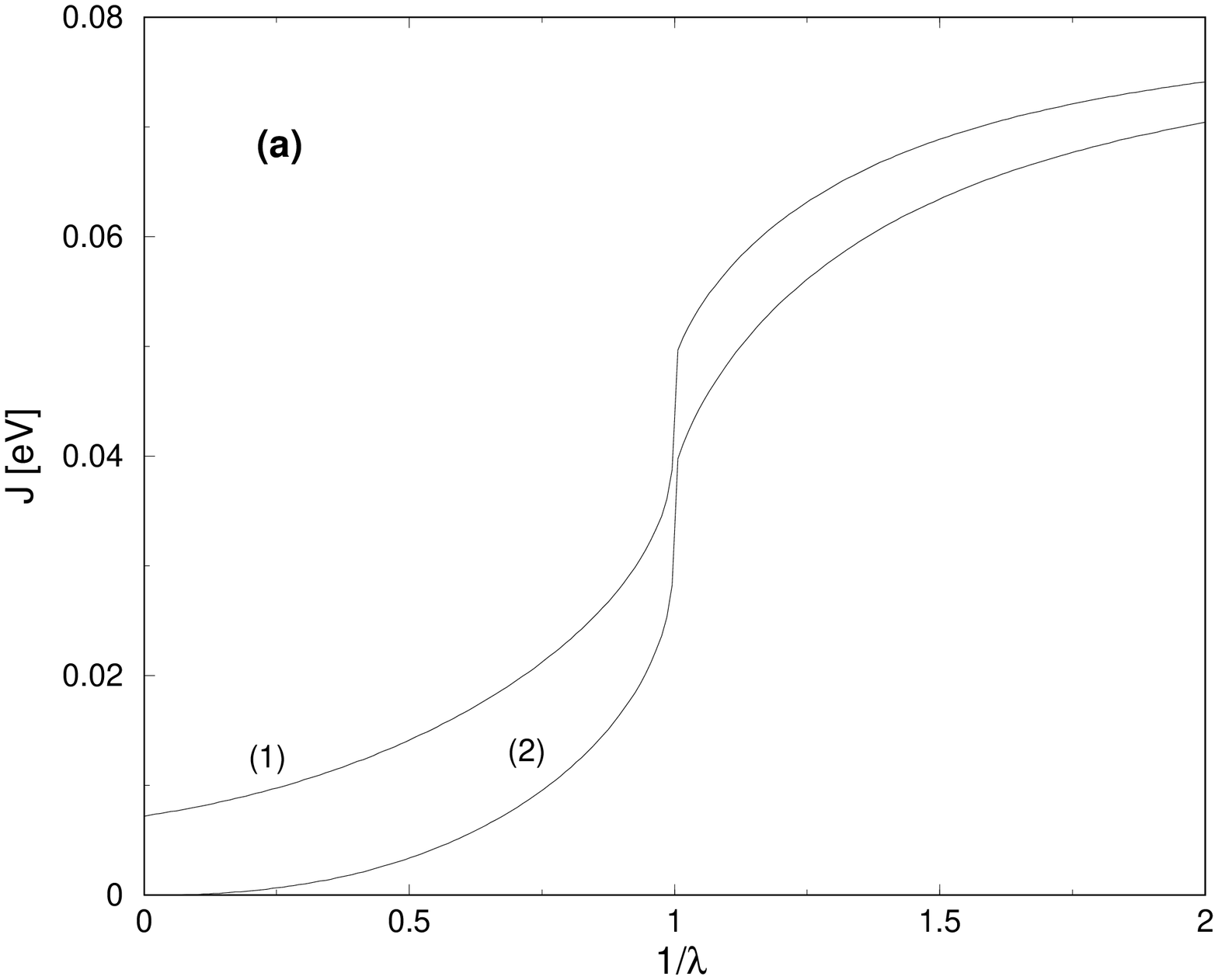,height=8cm,width=8cm,angle=-90}}
\centerline{\psfig{figure=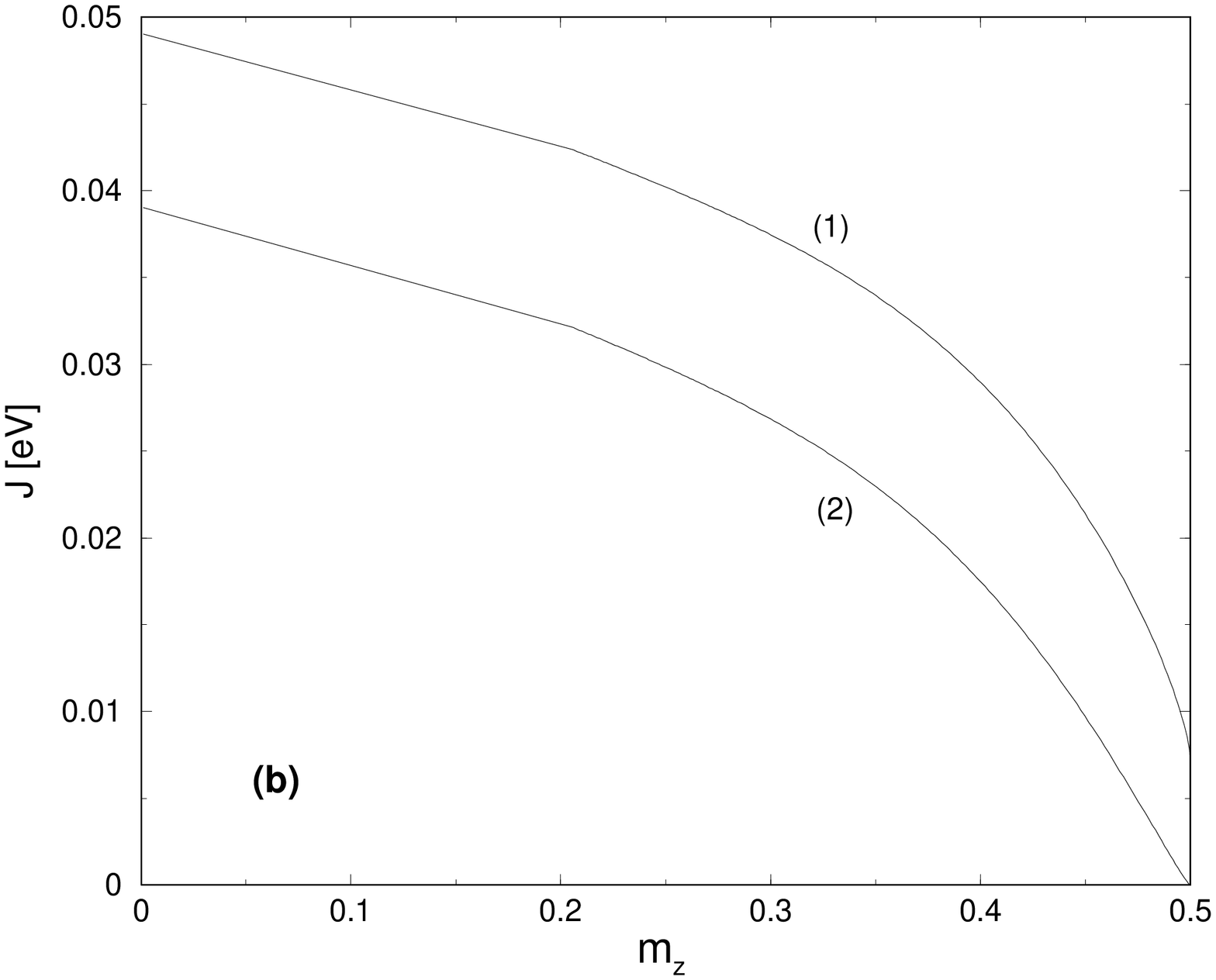,height=8cm,width=8cm,angle=-90}}
\vspace{1cm}
\caption{
Variation of the total exchange integral
$J=J^{(s)} + J^{(\tau)}$ in the wide range of the varying parameter
 $\lambda^{-1}$=$2t_a/V_b$ (a) and in the ordered state as function of the 
order parameter $m_z$ (b). The curves (1) and (2) correspond to two sets of
the hopping amplitudes with $t_d=t_b> 0$ and $t_d= 0$ 
respectively (see Sec. V).}
\end{figure}

\end{document}